\documentclass[%
 aip,
 amsmath,amssymb,
 reprint,%
]{revtex4-1}

\usepackage{graphicx}%
\usepackage{subcaption}
\usepackage[font=small,labelfont=bf,
   justification=justified,
   format=plain]{caption}
\usepackage{multirow}
\usepackage{dcolumn}%
\usepackage{bm}%

\usepackage[utf8]{inputenc}
\usepackage[T1]{fontenc}
\usepackage{mathptmx}
\usepackage{etoolbox}

\makeatletter
\def\@email#1#2{%
 \endgroup
 \patchcmd{\titleblock@produce}
  {\frontmatter@RRAPformat}
  {\frontmatter@RRAPformat{\produce@RRAP{*#1\href{mailto:#2}{#2}}}\frontmatter@RRAPformat}
  {}{}
}%
\makeatother
\begin{document}

\title{Boosting ensemble refinement with transferable force field corrections: synergistic optimization for molecular simulations}
\author{Ivan Gilardoni}
\author{Thorben Fröhlking}
\altaffiliation{ Current address: Université de Genève, Switzerland}%
\author{Giovanni Bussi}%

\affiliation{
$^1$Scuola Internazionale Superiore di Studi Avanzati, via Bonomea 265, 34136 Trieste, Italy\\
}%

\date{\today}%

\begin{abstract}
A novel method combining the ensemble refinement by maximum entropy principle and the force field fitting approach is presented. 
Its formulation allows to continuously interpolate in between these two methods, which can thus be interpreted as two limiting cases. 
A cross-validation procedure enables to correctly assess the relative weight of both of them, distinguishing scenarios where the combined approach is meaningful from those in which either ensemble refinement or force field fitting separately prevails. The efficacy of their combination is examined for a realistic case study of RNA oligomers. Within the new scheme, molecular dynamics simulations are integrated with experimental data provided by nuclear-magnetic-resonance measures. We show that force field corrections are in general superior when applied to the appropriate force field terms, but are automatically discarded by the method when applied to inappropriate force field terms.
\end{abstract}

\maketitle

\begin{quotation}
Abstract.
\end{quotation}

Molecular dynamics (MD) simulations play a crucial role in resolving the underlying conformational dynamics of molecular systems \cite{hollingsworth2018molecular}. However, their capability to reproduce and predict dynamics in agreement with experiments is limited by the statistical significance of the sampled trajectory and the accuracy of the force field model. While the first issue can be addressed by using enhanced sampling techniques \cite{henin2022enhanced}, the second one can be faced by suitable integration of MD simulations and experimental data \cite{bottaro2018biophysical}. To this aim, %
two main philosophies for experiment-based refinement were proposed in the literature \cite{orioli2020learn}. The first one is the so-called ensemble refinement (ER) approach \cite{pitera2012use,beauchamp2014bayesian,hummer2015bayesian,brookes2016experimental,kofinger2019efficient,bottaro2020integrating,medeiros2021comparison}. Developed from the maximum entropy principle, this technique selects the ensemble which best describes the experimental measures and is, at the same time, as close as possible to the initially hypothesized one. %
In doing so, ER is agnostic with respect to the knowledge of the force field parametrization: the functional form of the corrections carried to the initial ensemble only depends on the selected observables \cite{pitera2012use,cesari2018using} and thus the corrections are not transferable to different systems.
The second philosophy is force field refinement (FFR)
\cite{norgaard2008experimental,li2011iterative,wang2012systematic,wang2014building,cesari2016combining,cesari2019fitting,kofinger2021empirical,frohlking2022automatic}. Based on a reasonable guess of the %
force-field correction terms, their optimal coefficients are determined by minimizing a loss function that includes the discrepancy from experimental data and, in modern implementations, a regularization term \cite{wang2012systematic,wang2014building,cesari2016combining,cesari2019fitting,kofinger2021empirical,frohlking2022automatic} that penalizes moving away from the initial force field, in a Bayesian line of thought. This approach enables one to encode prior information about the reliability of a given force field term, by choosing which specific term should be refined, and makes the resulting corrections transferable to other systems \cite{frohlking2020toward}. However, adding the same force-field correction terms to all the copies of a given molecules or residue could be over-limiting and not able to capture further relevant differences among them. Indeed, the functional form of the force-field might be limited and intrinsically unable to reproduce experimental data.
These two categories of methods have been traditionally derived in a different manner. Only recently, a formulation of the FFR approach that formally relates to the maximum entropy principle has been proposed \cite{kofinger2021empirical}. Importantly, methods of the two classes have been so far used in a disjoint fashion. The user is thus expected to decide based on experience if transferable or non-transferable corrections are performing best for a given system.

In this Letter, we introduce a procedure to seamlessly combine the ER method with FFR. This allows to preserve the flexibility of ER while at the same time ensuring the transferability of the resulting force-field corrections to different molecules as in FFR. The procedure is here applied to the refinement of conformational ensembles of RNA oligomers, for which nuclear-magnetic-resonance (NMR) experimental data are available, %
but can be applied to reweight conformational ensembles of arbitrary systems for which solution data are available.
In a nutshell, the method works as follows. In traditional FFR approaches, the original ensemble is reweighted to include force-field corrections resulting in a new ensemble, which is then compared with experiment. Corrections are chosen so as to maximise the agreement. Here, before comparing with experiment, we perform an additional ER step, which fine tunes the resulting weights. The former step is expected to take into account any transferable contribution, and to leverage on the knowledge of which force-field terms might benefit a refinement. The latter step makes sure the final ensemble averages agrees with experiment.
The combination of the ER and FFR approaches is controlled by two hyper parameters ($\alpha$ and $\beta$). We first show the behavior of this approach on a toy model. Then, we use the method to derive ensembles and force field corrections for RNA oligomers. For the latter case, we show how a carefully performed cross-validation procedure is necessary to tune the hyper parameters. In our tests, we intentionally investigate the case where inappropriate force-field corrections are attempted, showing that our cross-validation procedure can detect this issue and automatically switch off the FFR step. %

\begin{figure}
\centering
\includegraphics[width=0.55\textwidth]{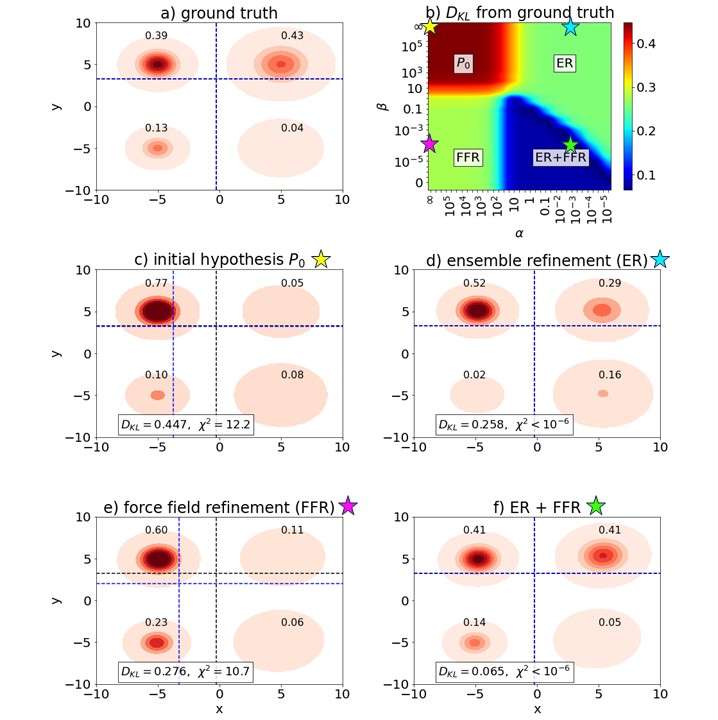}%
\captionsetup{justification=raggedright, singlelinecheck=off} %
\caption{Results on a toy model. (a) Ground truth distribution.  Populations are also reported for each peak. (c) Initially assumed distribution. (d) Ensemble refinement leads to an ensemble closer, but still not identical, to the ground truth.  %
(e) An attentive choice of the force field correction term could result in a different refinement (FFR), as good as the previous one (compare the Kullback-Leibler divergences from the ground truth, indicated as $D_{KL}$), with the benefit to be transferable. If this correction is not sufficient, (f) further flexibility can be included through the combined approach. In panels a,c,d,e,f, black dashed lines report the averages computed using the ground truth, whereas blue dashed lines report the averages computed using the refined ensemble. (b) Hyperparameters scan. The hyperparameters values prove to be crucial for a proper balancing of the ER and FFR contributions.
In panel (b), the values of the hyperparameters used to generate the ensembles in panels (c,d,e,f) are indicated with a star of a matching color.
}
\label{fig:toy_model}
\end{figure}

\textit{Proof of concept.}
To check the validity of the combined refinement, we set up a simple toy model which consists of a two-dimensional probability distribution with four peaks (see Figure \ref{fig:toy_model}).
The initial hypothesis sets most probability in the top-left peak ($x<0$ and $y>0$), while the ground truth probability is distributed also on the top-right peak ($x>0$ and $y>0$). So, the average value of the $x$ observable is underestimated by the initial hypothesis, whereas the average value of the $y$ observable is approximately correct. We then correct the initial hypothesis purely based on the value of the observed averages of $x$ and $y$. Ensemble refinement shifts the probability from the two peaks at $x<0$ to those at $x>0$, perfectly matching the observed averages. According to the maximum entropy principle, this is the minimal correction to the prior ensemble that allows matching the observed averages.
The resulting ensemble is closer to the ground truth one \cite{dannenhoffer2016direct}, but still not identical.
We then assume that a physical knowledge of the system suggests the top-right and bottom-left peaks to be coupled, leading to a specific functional form for the force-field correction. By performing a force-field refinement with this additional information, the observed averages are not exactly matched, but the obtained ensemble is also getting closer to the ground truth. Including further flexibility through the combined approach introduced in this work allows to optimally combine the information used in the force field refinement approach with the maximum entropy principle, resulting in better agreement with the ground truth ensemble than the one obtained applying any of the two methods separately. Figure \ref{fig:toy_model}b reports the distance from ground truth for the ensembles obtained using different possible values of $\alpha$ and $\beta$. The method interpolates between no correction to ensemble refinement, force-field refinement, and any combination of the two, as indicated in the figure. A similar figure reporting the discrepancy between the predicted and experimental observables ($\chi^2$) is reported in Fig. \ref{fig:chi2_toy_model}. 
A suitable choice of the two hyperparameters $\alpha,\,\beta$ is required to avoid overfitting. This is particularly relevant considering that experimental data are only known with a given uncertainty, which is not modeled in this toy model.

\begin{figure}
\centering
\includegraphics[width=0.45\textwidth]{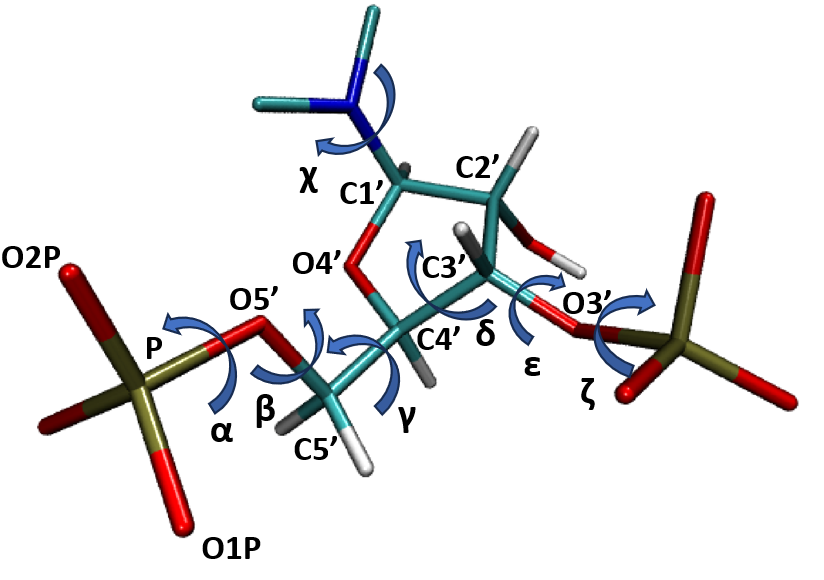}
\captionsetup{justification=raggedright, singlelinecheck=off} %
\caption{Sample RNA backbone structure, with standard atom names and dihedral angles indicated. The nucleobase is truncated so that only two carbon atoms from the cycle are shown.}
\label{fig:dihedrals}
\end{figure}

\begin{figure}
    \centering
    \captionsetup{justification=raggedright, singlelinecheck=off} %
    \begin{subfigure}[b]{0.21\textwidth} %
        \centering
        \caption[]{{\small RNA oligomers}} %
        \includegraphics[width=\textwidth]{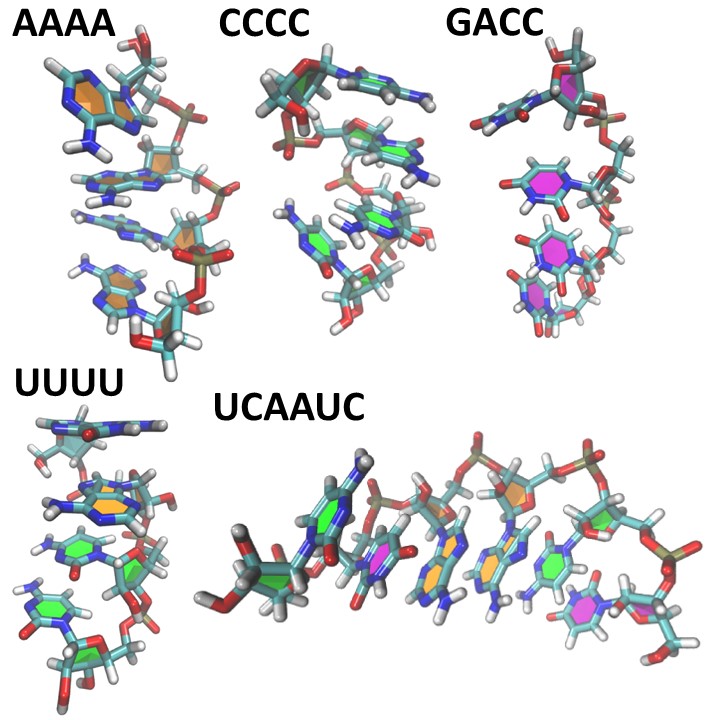}
           
        \label{fig: RNA molecules}
    \end{subfigure}
    \hfill
    \begin{subfigure}[b]{0.24\textwidth}  %
        \centering 
        \caption[]{{\small $\chi$ angles correction}}   %
        \includegraphics[width=\textwidth]{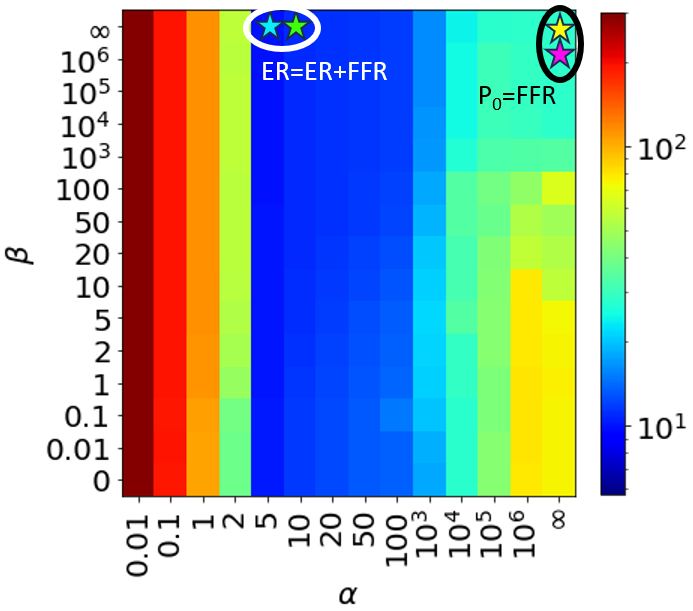}
          
    \end{subfigure}
    \vskip\baselineskip
    \begin{subfigure}[b]{0.23\textwidth}   %
        \centering 
        \caption[]{{\small $\alpha$ angles correction}}   
        \includegraphics[width=\textwidth]{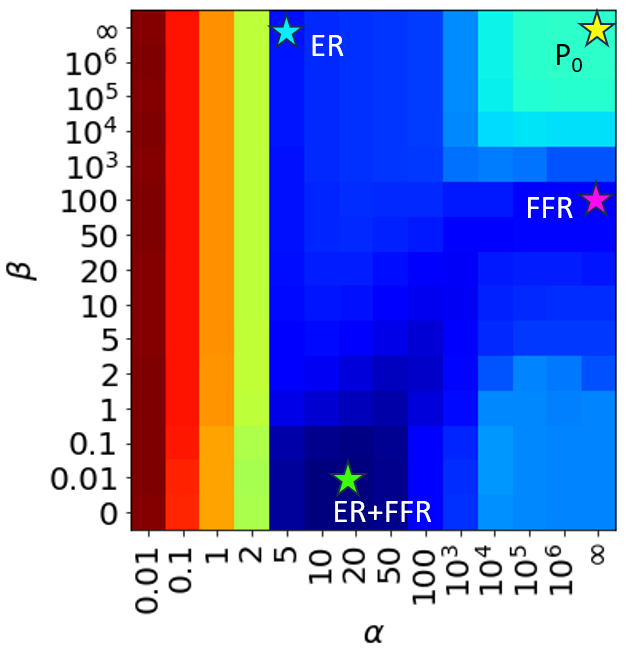}
         
    \end{subfigure}
    \hfill
    \begin{subfigure}[b]{0.23\textwidth}   %
        \centering
        \caption[]{{\small $\alpha,\,\zeta$ angles correction}} 
        \includegraphics[width=\textwidth]{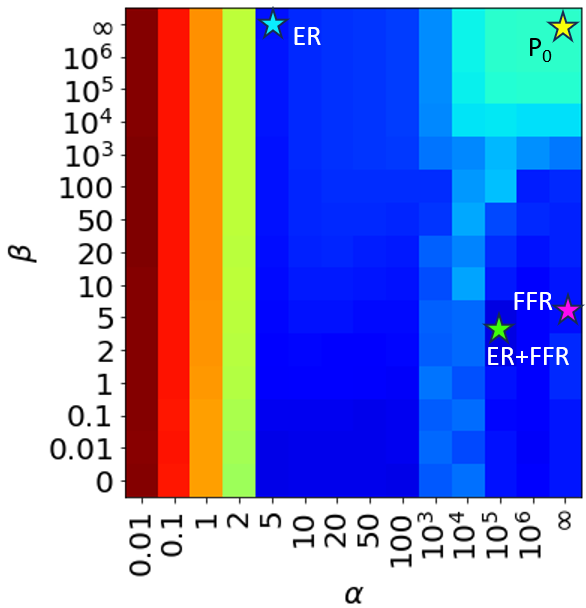}
        
    \end{subfigure}
    \caption[ case study ]
    {\small (a) Case study: RNA oligomers (4 tetramers and 1 hexamer). (b,c,d) Reduced $\chi^2$ on validating observables using three different functional forms for the force field refinement step (cross validation averages). In particular: (b) the correction on $\chi$ angles is fruitless and the contribution of (non-transferable) ensemble refinement is essential; (c) the correction on $\alpha$ angles is profitable, however adding more flexibility results in a better agreement with experimental values; (d) the correction on $\alpha,\,\zeta$ angles alone is enough and the inclusion of further flexibility is not necessary.}
    \label{fig:redchi2_validation}
\end{figure}

\textit{Application to real systems, including cross validation.}
We then test the method on a set of RNA oligomers for which simulations were previously reported \cite{frohlking2023simultaneous}, using the same experimental data set that was used in Ref. \cite{frohlking2023simultaneous}. Experimental data are taken from Refs.~\cite{condon2015stacking,tubbs2013nuclear,yildirim2011benchmarking,zhao2020nuclear,zhao2022nuclear}.
We perform the minimization of the loss function obtained combining all the oligomers in the training set (AAAA, CCCC, GACC, UUUU, UCAAUC -- Fig. \ref{fig:redchi2_validation}a) with a scan in the space of the hyperparameters $\alpha$ and $\beta$. The limiting cases of ensemble refinement (scan on the hyperparameter $\alpha$ at $\beta=\infty$) and force field fitting (scan on the hyperparameter $\beta$ at $\alpha=\infty$) are included, as well as the initially assumed ensemble, corresponding to $\alpha=\beta=\infty$ (i.e., no refinement). %
For each considered value of the hyperparameters $\alpha,\,\beta$, we take 20 random choices (seeds) of $70\%$ frames (of "demuxed" -- continuous -- trajectories, from replica exchange simulations, with the exception of UCAAUC for which a corrupted trajectory made it impossible to generate the continuous trajectories) and $70\%$ observables to be used as a training set (same choices for all the sampled hyperparameters), implementing a bootstrap strategy \cite{efron1994introduction}. The remaining observables are used to evaluate the reduced $\chi^2$ on the full trajectory, i.e., including also frames which were employed in training (validating step). The achievement of the minimizations is reflected by the increasing of the minimum value of the loss function at given seeds when $\alpha$ or $\beta$ are increased. 
Given the way trajectories are bootstrapped, the reduced $\chi^2$ on training observables and the one on validating observables coincide within their statistical error at $\alpha=\beta=\infty$, since it corresponds to the original ensemble.\par

The two hyper parameters ($\alpha$ and $\beta$) control the flexibility of the fitting. By decreasing one or both, the reliability given to the original assumptions on the force field is reduced in favour of the confidence on experimental measures. In other words, low values of the hyperparameters correspond to high flexibility in the correction to the ensemble, which means strong ability to fit the (training) experimental values. This is particularly true for the hyperparameter $\alpha$, corresponding to the ensemble refinement direction. For $\beta$ (force field refinement direction), the flexibility is instead intrinsically limited by the constrained functional form of the force field correction. Whereas a limited flexibility can provide a (physically meaningful) improvement in the description of the molecules, an uncontrolled flexibility may lead to overfit the data, disregarding their intrinsic experimental error. Cross validation is all about assessing the appropriate importance of this flexibility, which in the method here proposed plays on two different directions: the non-transferable ensemble refinement and the force field fitting ones. This task is performed by evaluating the error (the reduced $\chi^2$) on left-out observables, namely those which are not used in this  training step to determine the optimal ensemble.%

Whereas the $\chi^2_{red}$ computed on training data is decreasing when decreasing the values of the two hyperparameters, the $\chi^2_{red}$ computed on validating observables (Fig. \ref{fig:redchi2_validation} b,c,d) does so only over a certain range of the hyperparameters, signaling when under/over-fitting occurs. Firstly, we consider the case of traditional ensemble refinement, which in the combined method corresponds to $\beta=\infty$ (first row in the plots, independent on the selected force field correction; Fig. \ref{fig:redchi2_crossvalidation_separate}a). One can notice how, starting from the prior ensemble $P_0$ at $\alpha=\infty$, the error $\chi^2_{red}$ decreases with $\alpha$, up to a certain point, where it starts to increase. Such point of minimum (approximately at $\alpha\simeq5$, with $\chi^2\simeq 10$) marks the transition from the under-fitting to over-fitting scenarios, respectively. We then move forward to transferable corrections. It is instructive to consider three different functional forms for the force field refinement approach. All of them are given by linear combinations of sine and cosine of selected dihedral angle, respectively: $\chi$, O4$^\prime$--C1$^\prime$--N1$^\prime$--C2$^\prime$ for pyrimidines and O4$^\prime$--C1$^\prime$--N9$^\prime$--C4$^\prime$ for purines; $\alpha$, O3$^\prime_{i-1}$--P--O5$^\prime$--C5$^\prime$; and combined $\alpha,\,\zeta$, C3$^\prime$--O3$^\prime$--P$_{i+1}$--O5$^\prime_{i+1}$; in this last case we restricted to equal coefficients for sine terms and equal ones for cosine terms. 
Such force field corrections exhibit the three different behaviours which are expected when applying the combined ER+FFR method. The first attempted correction ($\chi$ angles, Fig.~\ref{fig:redchi2_validation}b and Fig. \ref{fig:redchi2_crossvalidation_separate}b) applied alone is fruitless, since the $\chi^2_{red}$ at $\alpha=\infty$, i.e. in the FFR regime, is larger than in the original ensemble $P_0$ for any choice of $\beta<\infty$. The contribution of (non-transferable) ensemble refinement is thus essential. Including both contributions, the minimum $\chi^2_{red}$ results at $\alpha=5,\beta=\infty$. Hence, this case corresponds to the extreme in which adding the contribution of the force field refinement does not improve the description with respect to ensemble refinement alone.
Even in this particularly difficult case, the ensemble refinement step is able to report reasonable cross-validation observables (see Table \ref{tab:hyperparameters}). The second attempted correction (on $\alpha$ angles) is profitable, as shown at $\alpha=\infty$ (FFR only, Fig. \ref{fig:redchi2_validation}c; see also Fig. \ref{fig:redchi2_crossvalidation_separate}c). However adding more flexibility with the combination of ER and FFR proves to be profitable with respect to the separate application of either the two methods since it results in a lower cross-validation error. Finally, the correction on $\alpha,\,\zeta$ angles alone is sufficient and the inclusion of further flexibility is not necessary (Fig. \ref{fig:redchi2_validation}d and Fig. \ref{fig:redchi2_crossvalidation_separate}d). This corresponds to the other extreme, in which FFR alone shows to be optimal.

\begin{table}
\captionsetup{justification=raggedright, singlelinecheck=off} %
\begin{ruledtabular}
\begin{tabular}{cccccccc}
force field correction & method & optimal $\alpha,\beta$ & $\chi^2_{red}$ \\ %
\hline
- & no reweighting & $\alpha=\infty,\,\beta=\infty$ & $28.32$\\
- & ER ($\beta=\infty$) & $\alpha=5$ & $10.10$ \\
$\chi$ angles & FFR ($\alpha=\infty$) & $\beta=\infty$ & $28.32$ \\
$\chi$ angles & ER+FFR & $\alpha=5,\,\beta=\infty$ & $10.10$ \\
$\alpha$ angles & FFR ($\alpha=\infty$) & $\beta=100$ & $9.33$ \\
$\alpha$ angles & ER+FFR & $\alpha=20,\,\beta=0.01$ & $5.70$ \\
$\alpha,\zeta$ angles & FFR ($\alpha=\infty$) & $\beta=5$ & $9.81$ \\
$\alpha,\zeta$ angles & ER+FFR & $\alpha=10^5,\,\beta=2$ & $8.02$ \\
\end{tabular}
\end{ruledtabular}
\caption{Results of cross validation on training molecules (see also Fig. \ref{fig:redchi2_validation}). We compare ER, FFR and their combination ER+FFR, with three different force field corrections. For each case, we report the optimal $\alpha,\,\beta$ hyperparameters and the minimum value of $\chi^2_{red}$ on validating observables (cross validation averages). In the first line we report the  $\chi^2_{red}$ before any corrections on the ensembles.}
\label{tab:hyperparameters}
\end{table}

\begin{figure}
    \centering
    \captionsetup{justification=raggedright, singlelinecheck=off} %
    \begin{subfigure}[b]{0.45\textwidth} %
        \centering
        \includegraphics[width=\textwidth]{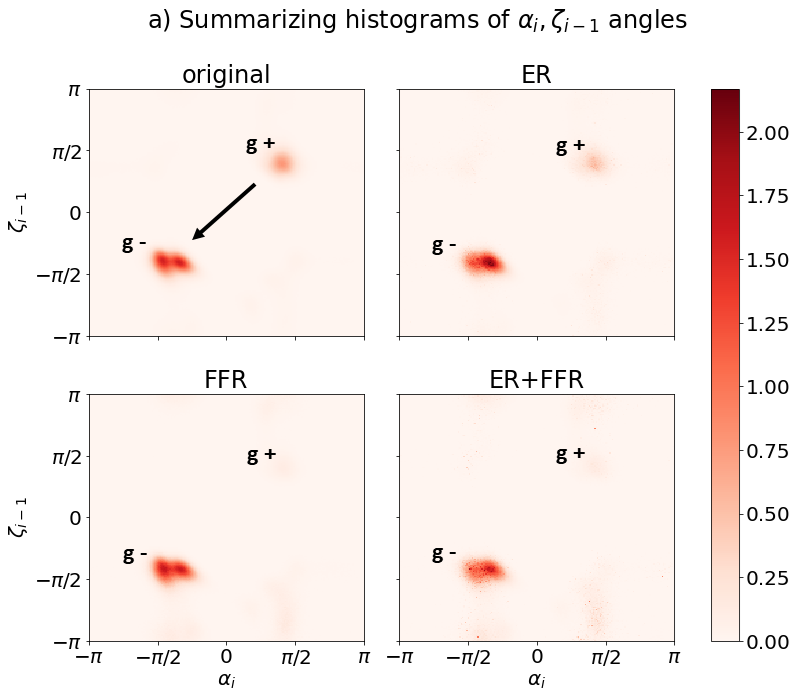}
           
        \label{fig: RNA molecules 2}
    \end{subfigure}
    \vskip\baselineskip
    \begin{subfigure}[b]{0.45\textwidth}   %
        \centering 
        \includegraphics[width=\textwidth]{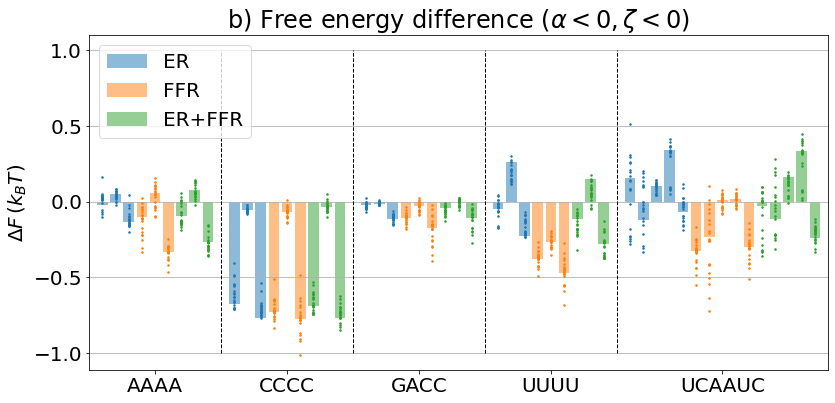}%
         
    \end{subfigure}
    \caption[ case study ]
    {(a) Summarizing histograms of $\alpha,\zeta$ angles before/after corrections on $\alpha$ dihedral angles, averaged over all the oligomers; (b) free energy difference for the $(\alpha,\zeta)\in(-\pi,0)\times(-\pi,0)$ region, for each molecule and phosphate (bars corresponding to minimization on the whole set of observables, dots corresponding to minimizations in cross validation).}
    \label{fig:histograms}
\end{figure}

\textit{Optimal force field corrections.}
Once the optimal values of the $\alpha,\beta$ hyper parameters have been determined through cross validation, we estimate the coefficients of the force field corrections by minimizing the loss function with such hyper parameters on the whole data set (without distinguishing between training and test set). The associated uncertainty is quantified by the standard deviation of the coefficients resulting from cross validation, as it is conventionally done in bootstrap analysis \cite{efron1994introduction}. The average values on the bootstrap samples are compatible with the values from whole minimization within the obtained uncertainties. The results are reported in Table \ref{tab:ffcorrections}. The correction on the glycosidic dihedral angles is null, since the optimal hyper parameters correspond to ensemble refinement only. This might imply that relevant force field corrections are not in the glycosidic bond, rather in the backbone structure. Indeed, for the other two cases (correction on $\alpha$ or $\alpha,\zeta$) the force field correction coefficients are significantly different from zero, with the general effect of disfavouring gauche$+$  conformations.

We then study how the force-field corrections on the $\alpha$ angles modify the distribution of $\alpha,\zeta$ dihedral angles, comparing the three different refinement methods described above. In these constructs, the phosphate group of the first nucleotide is absent (in agreement with experiments) so the $\alpha_i\,\zeta_{i-1}$ dihedral angles encompass the i-th phosphate group, with $i=2...N$. In Fig. \ref{fig:histograms}a) we show the overall histogram of the $\alpha_i,\zeta_{i-1}$ distributions, averaged over molecules and positions, which highlights two dominant peaks in the original ensembles, corresponding to gauche$+$ (g$+$) and gauche$-$ (g$-$). The population in these two peaks is modified when introducing ensemble and/or force field corrections, with a general increase in the $\alpha_{i}(g-),\zeta_{i-1}(g-)$ region, to the disadvantage of the $\alpha_{i}(g+),\zeta_{i-1}(g+)$ area. This is in agreement with previous studies \cite{gil2016empirical,bottaro2016free,chen2022rna}. To better visualize these variations, Fig.~\ref{fig:histograms}b reports the free energy differences $\Delta F$ associated to the $\alpha_{i}(g-),\zeta_{i-1}(g-)$ region, separately for each oligomer and phosphate group. We notice how, for AAAA, CCCC and GACC tetramers, the refinements have a significant impact only on the $\alpha_{i},\zeta_{i-1}$ angles corresponding to the first and last phosphate, while the population for the middle phosphate is almost unchanged. This can be explained on the basis that the employed force-field is well-suited for long RNA molecules, for which the first and last phosphate constitute a small fraction of the whole molecule, so that \emph{ad hoc} correction at the termini might be convenient\cite{mlynsky2020fine}. In particular, both the FFR and ER+FFR corrections go in the same direction as ER, favouring $\alpha(g-),\zeta(g-)$ angles as expected above. Results for the UUUU tetramer instead show significant free energy differences also for the intermediate phosphate. Here, ER and FFR suggest opposite corrections, with the former disfavouring $\alpha_3(g-),\,\zeta_2(g-)$.

Also for the UCAAUC hexamer, the $\alpha_i,\zeta_{i-1}$ dihedral angles corresponding to the last two phosphates tend to be modified by both the ER and ER+FFR methods, and left unchanged by FFR correction. Given the better performance of the ER+FFR approach in cross validation tests (see $\chi^2_{red}$ in Table \ref{tab:hyperparameters}), we argue that the ensembles obtained with the ER+FFR approach are more reliable than those obtained using the ER or the FFR approach alone.

It is also instructive to monitor the effect of the corrections to the least visited region in the $\alpha$,$\zeta$ domain (see Fig.~\ref{fig:few_visited}). This test highlights the difficulty of performing cross-validation tests for poorly populated region, for which modified weights might have a minor impact in both the discrepancy with
respect to experiment and the distance from the prior distribution.

\begin{table}
\centering
\captionsetup{justification=raggedright, singlelinecheck=off} %
\begin{ruledtabular}
\begin{tabular}{rcc}
ff correction / method & $\phi_1$ & $\phi_2$ \\ %
\hline
$\chi$ angles & $\phi_1\,(\sin\chi)$ & $\phi_2\,(\cos\chi)$ \\
FFR ($\alpha=\infty,\beta=\infty$) & $0$ & $0$ \\
ER+FFR ($\alpha=5,\beta=\infty$) & $0$ & $0$ \\
\hline
$\alpha$ angles & $\phi_1\,(\sin\alpha)$ & $\phi_2\,(\cos\alpha)$ \\
FFR ($\alpha=\infty,\beta=100$) & $0.91 \pm\, 0.16$ & $1.67 \pm\, 0.85$ \\
ER+FFR ($\alpha=20,\beta=0.01$) & $0.51 \pm\, 0.08$ & $1.65 \pm\, 0.36$ \\
\hline
$\alpha,\zeta$ angles & $\phi_1\,(\sin)$ & $\phi_2\,(\cos)$ \\
FFR ($\alpha=\infty,\beta=5$) & $3.0 \pm\, 1.5$ & $4.0 \pm \,2.0$ \\
ER+FFR ($\alpha=10^5,\beta=2$) & $3.0 \pm\, 1.6$ & $4.0 \pm \, 1.8$ \\
\end{tabular}
\end{ruledtabular}
\caption{Force-field correction coefficients (measure unit: $k_B T=2.49\,kJ/$mol). We report the coefficients resulting from the minimization on the whole data set together with their uncertainty. For the correction on $\chi$ angles, as shown in Fig. \ref{fig:redchi2_validation}, FFR provides no correction and optimal ER+FFR corresponds to ER only, hence no transferable corrections. The functional form for the force field corrections is $V(\chi)=\phi_1 \sin\chi+\phi_2\cos\chi$, $V(\alpha)=\phi_1\sin\alpha+\phi_2\cos\alpha$ and $V(\alpha,\zeta)=\phi_1(\sin\alpha+\sin\zeta)+\phi_2(\cos\alpha+\cos\zeta)$ respectively; sum over all the specified dihedral angles present in the molecule is implicit.}%
\label{tab:ffcorrections}
\end{table}

\textit{Testing the force field corrections on left-out molecules.}
Finally, we compare the performance of the two force fields that we obtained with the FFR and ER+FFR procedure when transferred to new molecules, not considered in training, namely the CAAU and UCUCGU oligomers. To this aim, we employ the optimal coefficients of the force field corrections that were reported in Table \ref{tab:ffcorrections} to reweight the CAAU and UCUCGU ensembles. We do so without performing a new ensemble refinement procedure.
The corresponding $\chi^2_{red}$ is then evaluated on the whole set of observables (see Table \ref{tab:validating_mols}).
These two molecules are quite different and respond differently to the correction fitted on the training set. Specifically, the original ensemble of CAAU has a very large $\chi^2$ which is dramatically decreased by the force field corrections. The larger the penalty on the gauche+ (g$+$) rotamers, the better the agreement with experiment. As a consequence, the correction applied on the  $\alpha,\zeta$ angles results in the best agreement with experiment. Conversely, the UCUCGU hexamer has a moderate $\chi^2$. Interestingly, the force field corrections obtained on the training set are not capable on improving the agreement of the corresponding ensemble with experiment. In this case, all corrections lead to some degree of overfitting. The mixed FFR+ER approach on the $\alpha$ angle, which leads to more conservative force field corrections, results in smaller overfitting and in a $\chi^2$ comparable to the one obtained with the original force field.

\begin{table}
\centering
\begin{tabular}{rccc}
\toprule
 & CAAU & UCUCGU & both \\ %
\hline
no reweighting & 243.8 & 14.1 & 211.3 \\
\hline
$\alpha$ angles correction &  &  \\
FFR & 24.8 & 16.4 & 23.6 \\
ER+FFR & 57.5 & 14.3 & 51.4 \\
\hline
$\alpha,\zeta$ angles correction &  &  \\
FFR & 11.7 & 23.8 & 13.4 \\
ER+FFR & 11.6 & 23.7 & 13.3 \\
\toprule
\end{tabular}
\caption{Reduced $\chi^2$ on validating molecules, based on the force-field corrections introduced above (coefficients resulting from minimization of the training molecules on the whole data set).}%
\label{tab:validating_mols}
\end{table}

In summary, this study reports a strategy to boost the efficiency of ensemble refinement methods by preceding them with a knowledge-based force field refinement step. This combined method allows also to obtain force-field corrections which can then be transferred to different systems, and can outperform normal ensemble refinement in cross validation tests. Differently from force field refinement by itself, these corrections are derived taking explicitly into account the fact that they might be unable to completely refine the different ensembles. This specificity on the system is guaranteed by the ensemble refinement term.
Whereas the force field refinement and ensemble refinement methods have been separately applied in several works, we are not aware of any attempt made to combine their strength points in a single approach. We employ the proposed method to a realistic case study of RNA oligomers. Despite the apparent simplicity of these small RNA molecules, current force-fields are still limited in correctly generating structural ensembles, which therefore can be used to improve molecular potentials. We apply a robust cross validation protocol to select the suitable values of the two hyper parameters $\alpha,\beta$ and then analyze the predictions about both training and validating oligomers.
The scripts used to perform the refinements discussed in this work can be found at \url{https://github.com/bussilab/force-field-ensemble-refinement}.
The analyzed time series can be found at \url{https://doi.org/10.5281/zenodo.10185005}.

In order to minimize the loss function we employ a reweighting of the ensembles approach. This might be a weakness of the proposed approach, common to all reweighting methods, due to its potentially low statistical efficiency\cite{rangan2018determination}. This concept can be quantified as the effective number of frames, computed through the Kish sample size or analogously the relative entropy. While for the examined oligomers the effective number of frames still remains a significant fraction of the whole amount (see Fig.~\ref{fig:kish_ratio}), for more complex systems, it might not be the case. Performing new MD simulations during the minimization, for instance whenever the resulting ensembles move too far from the initial ones, would lead to greater statistical robustness. Also, on-the-fly restraining could be performed \cite{cesari2016combining,cavalli2013molecular,white2014efficient,hummer2015bayesian,bonomi2016metainference}. A second, related issue arises from regions of the conformational space with limited or no sampling. Due to the way our cross validation procedure is performed, left-out portions of the initial trajectory are used to test for overfitting. However, if samples from a region are never observed in the initial trajectory, it is impossible to use reweighting to predict which will be the effect of the correction on samples from that region. If the region has a very large energy, e.g. because it is sterically forbidden, any change to the potential energy function is irrelevant. But if the region can be sampled in a new simulation for the same or for a different system, overfitting issues will arise \cite{frohlking2020toward}. This is a well-known issue in force-field fitting strategies, where it is common to perform new simulations using the refined force-field parameters to test for these artifacts. It might also be an issue in ensemble refinement maximum entropy strategies, if consecutive simulations are performed including linear corrections to the energy function. This issue is instead not expected to be visible if no new simulations are performed and the resulting ensemble is simply reported.

Finally, the discrepancy may be due not only to incorrect structural ensembles but also to inaccurate forward models used to compute experimental observables from MD simulations. In the most extreme cases, the ensembles might be in perfect agreement with the ground truth, still having high $\chi^2_{red}$ due to wrong forward models (like, for example, the empirical coefficients of Karplus equations).
In a recent work, we have shown how to simultaneously optimize ensembles and forward models \cite{frohlking2023simultaneous}. This idea could be pushed further and, in combination with the ideas presented in this work, lead to a simultaneous optimization of force fields, ensembles, and forward models.

\appendix

\nocite{*}
\bibliography{aipsamp}%

\clearpage

\section{Supplementary Methods}
\subsection{Combining ensemble and force field refinements}
Let's consider a molecular system, whose conformation is described by the high-dimensional vector $\vec x$ (containing all the coordinates of the atoms in the molecule). At thermal equilibrium at temperature $T$, the conformations $\vec x$ are distributed following the (ground truth) canonical ensemble $P_{GT}(\vec x)\propto \exp[-\beta U_{GT}(\vec x)]$, with $\beta=1/k_B T$ the Boltzmann factor and $U_{GT}(\vec x)$ the physical force field. This underlying force field is only partially known, and its best estimate we have is resumed in the initial hypothesis $U_0(\vec x)$, corresponding to the canonical distribution $P_0(\vec x)$. To improve our knowledge of the system, we can employ a set of experimental data $g_{i,exp}$, affected by uncertainties $\sigma_{i,exp}$, which approximate the average values of the corresponding observables $g_i(\vec x)$ over the ground truth ensemble.\par%

Based on this starting point, two approaches were developed in the literature: the Bayesian ensemble refinement (BioEn \cite{pitera2012use,hummer2015bayesian}) and the force field refinement\cite{cesari2019fitting,kofinger2021empirical,frohlking2022automatic}. Both of them suggest to modify the initial ensemble $P_0$ by decreasing its discrepancy with the new experimental data $g_{i,exp}\pm\sigma_{i,exp}$. However, we cannot rely only on these measurements, rather we should take into account all the previously collected theoretical and experimental information, which have led to our best (a priori) estimate $P_0$. This can be achieved by including a penalization to moving away from the initial ensemble $P_0$, which is to be interpreted as the prior knowledge in a Bayesian line of thought.\par%
In particular, the BioEn method minimizes the loss function
\begin{equation}
\mathcal{L}[P]=\frac{1}{2}\chi^2[P]+\alpha D_{KL}[P|P_0]
\label{eqn:lossf_ER}
\end{equation}
where the first term is
\begin{equation}
\frac{1}{2}\chi^2[P]=\frac{1}{2}\sum_{i=1}^{N_{exp}}\Bigl(\frac{\langle g_i\rangle_P-g_{i,exp}}{\sigma_{i,exp}}\Bigr)^2,
\label{eqn:chi2}
\end{equation}
(quantifying the discrepancy with the experimental measures $g_{i,exp}\pm\sigma_{i,exp}$) and the second one is the Kullback-Leibler divergence (the opposite of the relative entropy) of the ensemble $P$ with respect to the initial hypothesis $P_0$
\begin{equation}
D_{KL}[P|P_0]=%
\sum_x P(x)\log \frac{P(x)}{P_0(x)}
\end{equation}
(quantifying the discrepancy from $P_0$). The $D_{KL}$ is multiplied by a hyperparameter $\alpha$ which regulates the relative confidence given to the initial hypothesis with respect to the experimental data. Low $\alpha$ values little penalize deviations from $P_0$, favouring greater agreement with experimental values $g_{i,exp}$. The contrary happens for high $\alpha$ values. Minimizing this loss function leads to a balance between trusting experimental measures and initial hypothesis: the $\chi^2$ is kept down provided that $P$ is not too far from the initial ensemble. Through analytical calculations, one can show the optimal ensemble $P_*$ is determined by the system of implicit equations
\begin{equation}
P_*(x)=\frac{1}{Z_\lambda} P_0(x)\exp\Bigl(-\sum_{i=1}^{N_{exp}}\lambda_i g_i(x)\Bigr)    
\end{equation}
with
\begin{equation}
\label{eqn:lambda}
\lambda_i=\frac{\langle g_i\rangle_{P_*}-g_{i,exp}}{\alpha\sigma_{i,exp}^2}
\end{equation}
and $Z_\lambda$ the normalization factor $Z_\lambda=\langle e^{-\vec\lambda\cdot\vec g(x)}\rangle_0$. Taking advantage of this formal result, we can restrict the numerical minimization to the subset of ensembles parametrized by $\lambda$ coefficients $P_\lambda$, now let free to vary. Substituting $P_\lambda$ to $P$ in \eqref{eqn:lossf_ER} and employing \eqref{eqn:lambda}, one can show that minimizing $\mathcal L$ is equivalent to minimizing a different function
\begin{equation}
\Gamma(\lambda)=\frac{1}{2}\alpha\sum_j\sigma_j^2\lambda_j^2+\log Z_\lambda+\vec\lambda\cdot \vec g_{exp},
\label{eqn:gamma}
\end{equation}
with the advantage of fewer computational effort (same point of minimum $\lambda^*$ and minimum value $\mathcal L(P_{\lambda^*})=-\alpha \Gamma(\lambda^*)$). In particular, the covariance matrix of the observables $\langle g_i g_j\rangle_\lambda-\langle g_i\rangle_\lambda\langle g_j\rangle_\lambda$, which appears in the derivatives $\partial \mathcal L/\partial\mathcal \lambda_i$, simplifies when moving to the derivatives of $\Gamma$ function\cite{cesari2016combining}.\par

The second method, force field refinement, is based on minimizing a similar loss function,
\begin{equation}
\label{eqn:FFR_lossf}
\mathcal{L}[P_\phi]=\frac{1}{2}\chi^2[P_\phi]+\beta R[P_\phi|P_0] 
\end{equation}
being $R$ a regularizing term, which can be chosen as $D_{KL}[P_\phi|P_0]$ for the reason as above, and $\beta$ a hyperparameter playing the same role as $\alpha$ in the previous framework. The key difference with BioEn method is that now the variational freedom of the possible ensembles is limited to the selected functional form of the force field corrections $U_\phi(x)$
\begin{equation}
\label{eqn:P_phi}
P_\phi(x)=\frac{1}{Z_\phi} P_0(x) \exp(-\beta U_\phi (x)),
\end{equation}
being $\phi$ the flexible coefficients in the correction terms to the potential. 
It is exactly for this reason that the corrections brought through this method can be transferred to different molecules, at the expense of a limited flexibility. 
In this work, we consider only correction terms linear in the coefficients $\phi$, being $U_\phi(x)=\sum_j \phi_j f_j(x)$.\par %

In order to add further (and system-specific) flexibility to the force field corrections, one could consider to further refine the ensemble $P_\phi$ before comparing it with experimental measures, combining ER with FFR in this way. %
This can be achieved by minimizing such a loss function
\begin{equation}
\label{eqn:combined_lossf}
\mathcal{L}[P,P_\phi]=\frac{1}{2}\chi^2[P]+\alpha D_{KL}[P|P_\phi]+\beta D_{KL}[P_\phi|P_0].
\end{equation}
where the two hyperparameters $\alpha,\,\beta$ regulate the reliability of the initial hypothesis. This corresponds to introducing ER as an intermediate step in between the FFR step and the comparison with experiments (compare eq.s \eqref{eqn:FFR_lossf} and \eqref{eqn:combined_lossf}). Such a method reduces to FFR only for $\alpha=\infty$, i.e., maximally penalizing flexible deviations from $P_\phi$, and to ER only for $\beta=\infty$ (i.e., no force field refinement). %
Repeating the same calculations as before, the optimal ensemble $P_*$ satisfies%
\begin{equation}
\label{eqn:P_lambdaphi}
P_*(x)=\frac{1}{Z_{\lambda\phi}} P_\phi(x) \exp\Bigl(-\sum_{i=1}^{N_{exp}}\lambda_i g_i(x)\Bigr),
\end{equation}
with same relation for $\vec\lambda$ as in \eqref{eqn:lambda} and $Z_{\lambda\phi}=\langle e^{-\vec\lambda\cdot\vec g(x)}\rangle_\phi$ the normalization factor. Again, this allows to employ the $\vec\lambda,\vec\phi$ parametrization in order to restrict the minimization of the loss function to the ensembles $P_{\lambda\phi}$.

\subsection{Interpretation of the hyperparameters}
The hyperparameter $\alpha$ regulates how much we consider the force field correction $U_\phi$ able to catch all the missing refinement, with respect to the experimental accuracy. Low values of $\alpha$ little oppose to deviations from $P_\phi$, provided that better (even small) agreement with experimental values $g_{i,exp}$ take place, at the contrary for high $\alpha$ values where the same deviation from $P_0$ have to be supported by huge decreasing in the $\chi^2$. The hyperparameter $\beta$ tunes our degree of reliability towards the prior potential $U_0$, regarding only the $U_\phi$ correction.\par

The following limiting cases appear:
\begin{itemize}
    \item $\alpha,\,\beta\rightarrow\infty$: we completely trust our initial hypothesis (both the force field model and its coefficients) $P=P_\phi=P_0$; the unique finite value of the loss function is the $\chi^2$ for the original ensemble;
    \item only $\alpha\rightarrow\infty$: we believe the force field correction $U_\phi$ to be able to catch all the missing information required to appropriately describe the molecules, and we only need to refine its coefficients; all the possible ensembles restrict to $P=P_\phi$, i.e., the combined approach reduces to FFR only;%
    \item only $\beta\rightarrow\infty$: we believe refining the force field correction $U_\phi$ does not improve the description of the systems (so we keep $P_\phi=P_0$), and further (possibly system-specific) flexibility is required; the combined approach reduces to ER only;%
    \item $\alpha,\,\beta\rightarrow0$: the initial assumptions are considered quite unreliable with respect to the accuracy of experimental values. Limiting cases where the hyperparameters are very small can lead to overfitting the experimental data, disregarding their intrinsic experimental uncertainty. Notice $\alpha=0$ (with finite $\beta$) uncouples the comparison with experiments from the force field refinement, making the combined approach meaningless. %
\end{itemize}

By choosing finite values of $\alpha,\,\beta$ we can optimize the performance of the combined approach. Such optimization of the hyperparameters is described in the Cross validation section.

\subsection{Calculations using reweighting}
In the above derivation we assumed to be able to compute averages over ensembles $P_0$, $P_{\lambda\phi}$ and $P_\phi$. By performing MD simulations, and assuming ergodicity of the sampled trajectories, we can replace averages over the configuration space with averages over frames (with normalized weight $w_t$) of the corresponding MD trajectories (possibly selected one every N, in order to decrease correlation between consecutive frames):
\begin{equation}
\langle \mathcal{O}\rangle_P=\sum_x \mathcal{O}(x) P(x)\simeq\sum_t \mathcal{O}_t w_t.
\end{equation}
However, this would require to perform MD simulations for each demanded ensemble, which is in practice unfeasible. To circumvent this issue, one can exploit the relationships of $P_{\lambda\phi}$ and $P_\phi$ to the initial ensemble $P_0$ (see eq.s \eqref{eqn:P_lambdaphi}, \eqref{eqn:P_phi}). In this way, once the initial ensemble $P_0$ has been accurately sampled through a MD simulation, the expectation values over $P_{\lambda\phi},\,P_\phi$ can be traced back to computing averages over $P_0$, in a reweighting procedure. %
In particular, averages over $P_\phi$ and $P_{\lambda\phi}$ can be computed respectively as
\begin{equation}
\langle \mathcal{O} \rangle_\phi=\frac{1}{Z_\phi}\sum_x \mathcal{O}(x) P_0(x) e^{-\beta U_\phi(x)}%
=\frac{\langle\mathcal{O}(x)e^{-\beta U_\phi(x)}\rangle_0}{\langle e^{-\beta U_\phi(x)}\rangle_0}   
\end{equation}
\begin{equation}
\begin{split}
\langle \mathcal{O}\rangle_{\lambda\phi}&=\frac{1}{Z_{\lambda\phi}}\frac{1}{Z_\phi}\sum_x \mathcal{O}(x) P_0(x) e^{-\beta U_\phi(x)}e^{-\vec\lambda\cdot\vec g(x)}\\
&=\frac{\langle \mathcal{O}(x)e^{-\vec\lambda\cdot \vec g(x)}e^{-\beta U_\phi(x)}\rangle_0}{\langle e^{-\vec\lambda\cdot \vec g(x)}e^{-\beta U_\phi(x)}\rangle_0}
\end{split}
\end{equation}
being $\langle ...\rangle_0$ the averages over the initial ensemble $P_0$.
Clearly, the sampling of $P_0$ is limited, so moving away from $P_0$ the fidelity of the reweighted average decreases. Such goodness can be quantified through the effective n. of frames $N_{eff}$ of the reweighted ensemble, which can be estimated by the so-called relative Kish sample size $K_{rel}$ as
\begin{equation}
N_{eff}/N=K_{rel}[\{w_t\}|\{w_{0t}\}]=\Bigl(\sum_t \frac{w_t^2}{w_{0t}}\Bigr)^{-1}
\end{equation}
or analogously by the exponential of the Kullback-Leibler divergence $e^{D_{KL}[P|P_0]}$. Both these two indicators measure the average relative ratio $w/w_0$, as arithmetic or geometric mean respectively.

\subsection{Generalization to multiple systems}
In order to fully exploit the potentiality of the presented method, it is convenient to consider more than one molecular system. In this way, one can achieve a better refinement of the force field parameters $\vec\phi$, which are common to different systems. Let's use the index $k$ to enumerate such molecules $k=1...N_{sys}$, with $P^k$ the corresponding ensembles and $\vec\lambda_k$ the system-specific ER coefficients. The loss function can be written as the sum of \eqref{eqn:combined_lossf} over different systems
\begin{equation}
\mathcal{L}\Bigl(\{\vec\lambda_k\},\vec\phi\Bigr)=\sum_{k=1}^{N_{sys}}\mathcal L[P^k_{\lambda_k\phi},P^k_\phi]
\label{eqn:lossf_complete}
\end{equation}
In this way, each system contributes to the $\chi^2$ with a weight proportional to the number of its experimental data. Notice that, in principle, each system could be included with separated hyper parameters $\alpha,\,\beta$, or with a separate weight for the $\chi^2$. In this work, we consider a homogeneous set of systems, so we do not exploit this possibility.

\subsection{Minimization strategy}
Firstly, let's focus on a single system. Minimizing the loss function $\mathcal L(\vec\lambda,\vec\phi)$ (given by \eqref{eqn:combined_lossf} with employing the $\vec\lambda$ parametrization) would require to compute the covariance matrix
\begin{equation}
C_{ij}(\lambda,\phi)=\langle g_i g_j \rangle_{\lambda\phi}-\langle g_i\rangle_{\lambda\phi} \langle g_j\rangle_{\lambda\phi},
\end{equation}
which appears in the $\vec\lambda$ derivatives
\begin{equation}
\frac{\partial \mathcal L}{\partial \lambda_i}=-\sum_j\Bigl(\frac{\langle g_j\rangle_{\lambda\phi}-g_{j,exp}}{\sigma_{j,exp}^2}-\alpha\lambda_j\Bigr)C_{ij}(\lambda,\phi)%
\label{eqn:der_lambda1}
\end{equation}

However, this is quite expensive when the number $M$ of available experimental data is high, since the number of independent components in the covariance matrix scales as $M^2$. To decrease this computational effort, we can translate such minimization problem to searching for a saddle point of a different function $\mathcal L'(\vec\lambda,\vec\phi)$. Even if the search for a saddle point is in general more challenging than minimization, the new function $\mathcal L'$ is considerably simpler than $\mathcal L$, resulting in significant improvement. Indeed, performing firstly the minimization over $\lambda$ at fixed $\phi$ and proceeding as in \eqref{eqn:gamma} (substituting \eqref{eqn:lambda} in \eqref{eqn:combined_lossf} parametrized by $\vec\lambda$), we get
\begin{equation}
\mathcal{L}'(\vec\lambda,\vec\phi)=-\alpha \Gamma_\phi(\lambda)+\beta D_{KL}[P_\phi|P_0]
\label{eqn:lossf_saddle}
\end{equation}
being
\begin{equation}
\Gamma_\phi(\lambda)=\frac{1}{2}\alpha\sum_j \sigma_j^2 \lambda_j^2+\log Z_{\lambda\phi}+\vec\lambda\cdot\vec g_{exp},
\end{equation}
which is the same function derived in the maximum relative entropy approach \cite{cesari2016combining}, except that now it is evaluated with respect to the reference ensemble $P_\phi$, rather than $P_0$. %
The derivatives w.r.t. $\vec\lambda$ considerably simplify to
\begin{equation}
\frac{\partial \mathcal L'}{\partial \lambda_j}=\alpha(\langle g_j\rangle_{\lambda\phi}-g_{j,exp}-\alpha\sigma_j^2\lambda_j).
\end{equation}

Assuming corrections to the force field which are linear in the coefficients $\vec\phi$, we can explicate $D_{KL}$ in \eqref{eqn:lossf_saddle} and the loss function $\mathcal L'$ can be written as
\begin{equation}
\mathcal L'(\lambda,\phi)=-\alpha\Gamma_\phi(\lambda)-\beta(\log Z_\phi+\vec\phi\cdot\langle\vec f\rangle_\phi),
\end{equation}
with derivatives w.r.t. $\vec\phi$
\begin{equation}
\frac{\partial \mathcal L'}{\partial \phi_k}=\alpha(\langle f_k\rangle_{\lambda\phi}-\langle f_k\rangle_{\phi})+\beta\vec\phi\cdot(\langle \vec f f_k\rangle_\phi-\langle \vec f\rangle_\phi\langle f_k\rangle_\phi).
\label{eqn:der_lambda2}
\end{equation}

Moving from $\mathcal L$ to $\mathcal L'$ flipped the minimum in the $\vec\lambda$ direction to a maximum, and so the global minimum of $\mathcal L$ has become a saddle point (also called mini-max, due to its characteristic property): 
\begin{equation}
(\lambda^*,\phi^*)=\arg\min_{\lambda,\phi}\mathcal L(\lambda,\phi)=\arg \min_\phi \max_\lambda \mathcal L'(\lambda,\phi).
\end{equation}
The search for the saddle point can be tackled by performing a nested minimization, with the inner one of $\Gamma_\phi(\lambda)$ over $\lambda$ and the outer over $\phi$ coefficients:
\begin{equation}
(\lambda^*,\phi^*)=\arg\min_\phi \Bigl(-\alpha \min_\lambda \Gamma_\phi(\lambda)+\beta D_{KL}[P_\phi|P_0]\Bigr).  
\end{equation}
Both the inner and outer minimizations can be performed with second-order methods, i.e. preconditioning the gradient with curvature information, by employing BFGS or L-BFGS-B methods (provided by the Python library Scipy).\par

Alternatively to the search for the saddle point, one could perform adaptive moves in the space of $\lambda$ coefficients, such that the covariance matrix (positive definite for independent observables $g_i$) which multiplies the derivatives' vector $\partial\mathcal L/\partial\lambda_j$ simplifies. Such moves correspond to iteratively normalize the observables $g_j(x)$ at each step. From a conceptual point of view, this corresponds to natural gradient descent\cite{amari1998natural} in the space of $P_\lambda$ probability distributions (the step length in each direction is proportional to the variation of the ensemble in that direction, measured through the relative entropy). However, this does not provide curvature information, required to perform second-order minimization (like BFGS), so we opted for the search for the saddle point (i.e., nested minimization).\par

For multiple systems, we proceed analogously, taking into account that, while for the $\vec\lambda$ coefficients each system separately contributes to the gradient, when computing the $\vec\phi$ components one has to sum the contributes each system gives to each force field coefficient $\vec\phi$.

\subsection{Cross validation}

To find suitable values for the hyperparameters, we perform a cross validation procedure. The whole data set is iteratively split into training and test set. This partition is made by randomly selecting $70\%$ frames and $70\%$ of observables as a training set for each molecular system. In the following, parallel tempering will be used as enhanced sampling technique in MD simulations, so we will opt to select frames by choosing approximately $70\%$ of demuxed trajectories. The remaining data, which corresponds to non-selected frames (all the observables) or selected frames and non-selected observables, will be used as a test set for that particular iteration of cross validation. For each of these partitions, a minimization of the loss function is performed on the training set, with a scan over different values of the hyperparameters. Once the optimal ensemble (at fixed hyperparameters) has been found, we evaluate the (reduced) chi squared $\chi^2_{red}$ for that particular ensemble using data from the test set. Starting from high values of the hyperparameters $\alpha,\,\beta$ and decreasing them, we expect a transition from the under-fitting case (when the initial assumptions are still dominant with respect to experiments) to the over-fitting scenario (when, on the contrary, the importance given to experimental measures is too strong). This transition will be marked by a minimum of the reduced chi squared on the test set. In this way, we have a practical procedure to determine the reliability of the initial model (both of the force field functional form and the initial coefficients of its correction terms). Once the optimal value of the hyperparameters has been found, one could determine the optimal ensemble by minimizing the loss function on the whole data set. This result can be eventually compared with the "average ensemble" resulting from the minimizations performed in cross validation at corresponding values of the hyperparameters. However, defining properly what average ensemble means is far from trivial, also because of different $\lambda$ coefficients, depending on the selected observables.

\subsection{Simulation details}
We apply the method introduced above to the refinement of RNA structural ensembles. We employ MD simulations performed in Ref. \cite{frohlking2023simultaneous} for a set of RNA tetramers with sequence AAAA, CAAU, CCCC, GACC, UUUU and for two hexamers with sequence UCAAUC and UCUCGU (fig. \ref{fig:redchi2_validation} a). In particular, the CAAU tetramer and UCUCGU hexamer are used as validating molecules, to test the transferability of the force field refinement determined by the other molecular systems. MD simulations were performed with the standard OL3 RNA force field \cite{cornell1996second,wang2000well,zgarbova2011refinement,steinbrecher2012revised} with the van der Waals modification of phosphate oxygens developed in ref. \cite{perez2007refinement}. Explicit solvent was used, with OPC\cite{izadi2014building} water model and KCl salt concentrations\cite{joung2008determination} corresponding to the experimental conditions to which MD simulations will be compared. To enhance sampling, parallel tempering was used with 24 replicas. The temperatures were chosen ranging from $275\,K$ to $400\,K$ and systems were simulated for $1\,\mu s$ per replica. Only the trajectory closest to room temperature $300\,K$ is employed in the following analysis. For more simulation details, we refer to Ref. \cite{frohlking2023simultaneous}.

\subsection{Experimental data}
All RNA oligomers introduced above can be compared to experimental studies providing nuclear-magnetic-resonance (NMR) data, which includes $^3J$ scalar couplings as well as observed NOEs (Nuclear Overhauser Effect signals) and unobserved NOEs (shortened to uNOEs).\par

The $^3J$ couplings are frequency differences (measured in Hz) due to spin-spin couplings between close nuclei mediated through chemical bonds. In this work, we employ three-bond couplings, whose magnitude provides information on the dihedral angles relating the coupling partners, in particular the $\beta$ ($H_5-C_5-O_5-P$), $\epsilon$ ($H_3-C_3-O_3-P$), $\gamma$ ($H_5-C_5-C_4-H_4$) dihedral angles of backbone and the $\nu$ dihedral angles of sugar ($H-C-C-H$). The experimental values, together with the corresponding estimated uncertainties, are provided by Refs. ~\cite{condon2015stacking,tubbs2013nuclear,yildirim2011benchmarking,zhao2020nuclear,zhao2022nuclear}. The relation between $^3J$ couplings and dihedral angles is described by the Karplus equations with empirical coefficients
\begin{equation}
^3J(\phi)=A\cos^2\phi-B\cos\phi+C
\end{equation}
where $\phi$ is the involved dihedral angle. These equations will be used as a forward model to back-calculate the scalar couplings from dihedral angles provided by MD simulations. Their coefficients were estimated in Refs. \cite{lankhorst1985carbon,davies1978conformations,condon2015stacking}.

The Nuclear Overhauser Effect (NOE) is the transfer of nuclear spin polarization from one $H$ nucleus (single proton) to a close one, when the second is saturated by radio-frequency irradiation. This effect %
results in a signal which is proportional to $1/r^6$ where $r$ is the distance between the two protons. Whenever the two protons are farther than a certain threshold, NMR experiments are not able to register any NOE signal, therefore we indicate this as unobserved NOE signal (uNOEs). Both NOEs and uNOEs experimental values are provided with associated uncertainties by Refs. \cite{condon2015stacking,tubbs2013nuclear,yildirim2011benchmarking,zhao2020nuclear,zhao2022nuclear} as distances $r$ (in \AA). However, they are compared with MD simulations in the $\chi^2$ as signals $1/r^6$ rather than distances. To evaluate the corresponding error, given the uncertainty $\delta r$, we take the average
\begin{equation}
\delta \Bigl(\frac{1}{r^6}\Bigr)=\frac{1}{2}\Bigl(\Big |\frac{1}{(r-\delta r)^6}-\frac{1}{r^6}\Big|+\Big|\frac{1}{(r+\delta r)^6}-\frac{1}{r^6}\Big|\Bigr).
\end{equation}
Notice that unobserved NOEs correspond to an upper threshold in the signal, so they only have upper uncertainty 
\begin{equation}
\delta \Bigl(\frac{1}{r^6}\Bigr)=\frac{1}{(r-\delta r)^6}-\frac{1}{r^6}.
\end{equation}

Hence, when computing the $\chi^2$ terms for the unobserved NOE observables $g(x)$, the expression \ref{eqn:chi2} is substituted by
\begin{equation}
\chi^2[P]=\Bigl[\max\Bigl(0,\frac{\langle g\rangle_P-g_{exp}}{\sigma}\Bigr)\Bigr]^2,
\end{equation}
corresponding to non-reported error if the average value from MD simulation is lower than the experimental threshold $g_{exp}$.

\section{Supplementary Results}

\subsection{Toy model}
\begin{figure}
\centering
\includegraphics[width=0.5\textwidth]{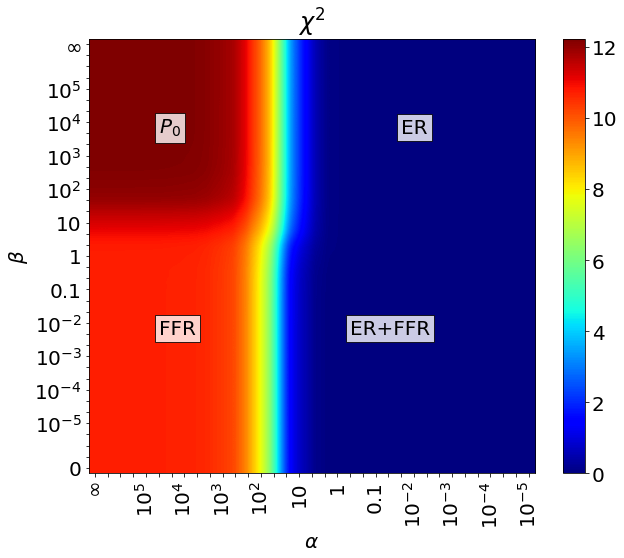}
\captionsetup{justification=raggedright, singlelinecheck=off} %
\caption{Results on a toy model: $\chi^2$ for the optimal ensemble as a function of $\alpha,\beta$ hyper parameters.}
\label{fig:chi2_toy_model}
\end{figure}

In Fig. \ref{fig:chi2_toy_model} we report the $\chi^2$ of the optimal ensemble as a function of $\alpha,\beta$ hyper parameters. Starting from the original ensemble $P_0$ (ground truth) at $\alpha=\beta=\infty$, which corresponds to $\chi^2\simeq 12.2$, and decreasing the value of one or both the two hyper parameters, the $\chi^2$ tends to decrease. However, this occurs remarkably different for $\alpha$ (ensemble refinement) or $\beta$ (force field refinement) directions: in the first case, turning down the hyper parameter $\alpha$ at given $\beta$, the $\chi^2$ decreases down to values close to $\chi^2=0$, while in the other case the "stiffness" of the force-field functional form stops the decrement at the limit value $\chi^2\simeq 10.7$. Starting from low values of $\beta$ and allowing for more flexibility of the ensemble ($\alpha$ direction), the $\chi^2$ decreases down to $\chi^2=0$, as for ER only. However, although the $\chi^2$ is almost the same, the Kullback-Leibler divergence from the ground truth is significantly different in the two regions ER and ER+FFR: including some knowledge about the force-field results in better refinement.

\subsection{RNA oligomers}

\subsubsection{Cross validation procedure}

\begin{figure*}
    \centering
    \captionsetup{justification=raggedright, singlelinecheck=off} %
    \begin{subfigure}[b]{0.4\textwidth} %
        \centering
        \caption[]{{\small Ensemble refinement}} %
        \includegraphics[width=\textwidth]{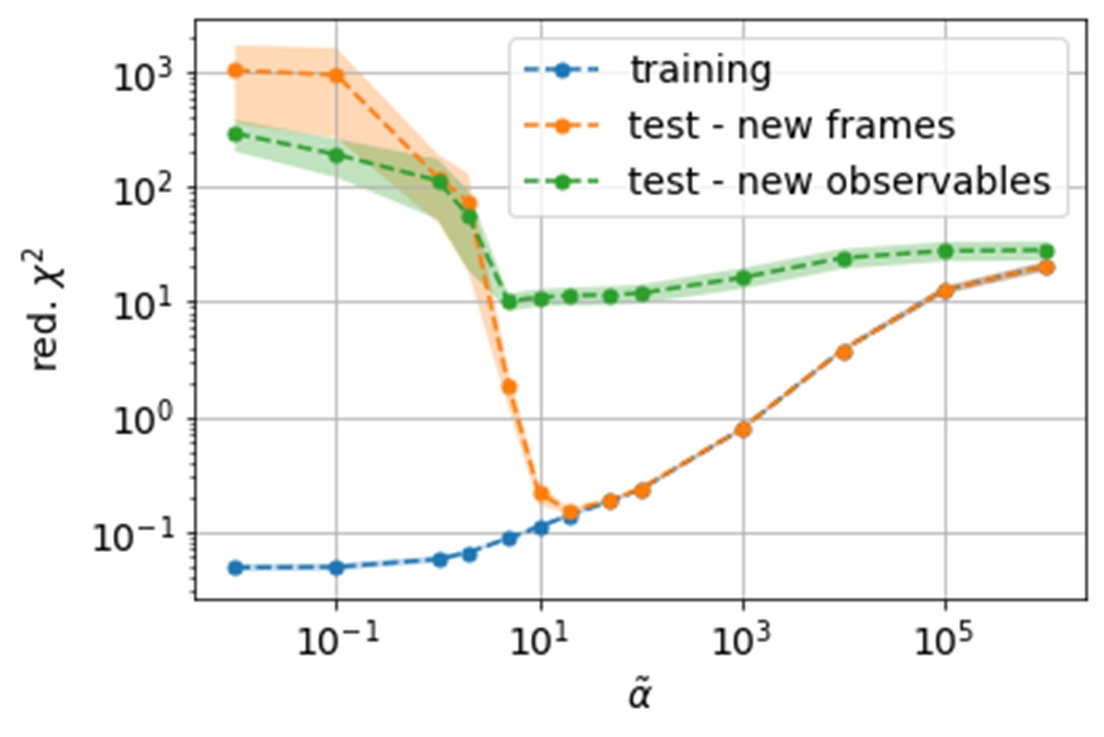}
           
        \label{fig: RNA molecules 3}
    \end{subfigure}
    \hfill
    \begin{subfigure}[b]{0.4\textwidth}
        \centering 
        \caption[]{{\small $\chi$ angles correction}}   %
        \includegraphics[width=\textwidth]{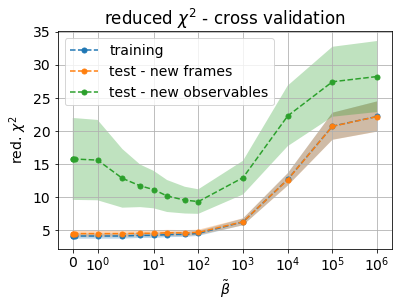}
          
    \end{subfigure}
    \vskip\baselineskip
    \begin{subfigure}[b]{0.4\textwidth}
        \centering 
        \caption[]{{\small $\alpha$ angles correction}}   
        \includegraphics[width=\textwidth]{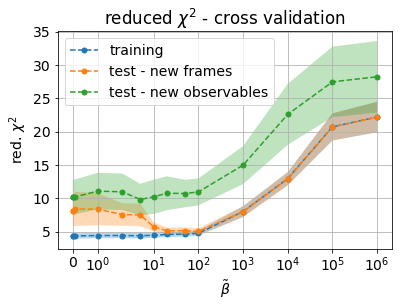}
         
    \end{subfigure}
    \hfill
    \begin{subfigure}[b]{0.4\textwidth}
        \centering
        \caption[]{{\small $\alpha,\,\zeta$ angles correction}} 
        \includegraphics[width=\textwidth]{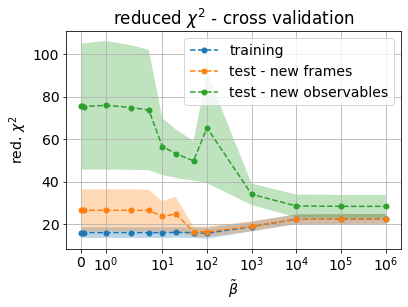}
        
    \end{subfigure}
    \caption[ case study ]
    {\small Reduced $\chi^2$ in cross validation for ensemble refinement and force field refinement separately applied.}
    \label{fig:redchi2_crossvalidation_separate}
\end{figure*}

In Fig. \ref{fig:redchi2_crossvalidation_separate} we report the reduced $\chi^2$ for ensemble refinement and force field refinement on RNA oligomers, separately applied. The force field corrections include the three choices previously described, i.e., the sinusoidal corrections on $\alpha$, $\alpha$ and $\zeta$, $\chi$ dihedral angles respectively. For the correction on $\alpha$ and $\zeta$ angles we fixed the coefficients of $\sin\alpha$, $\sin\zeta$ to be equal, the same for $\cos\alpha$, $\cos\zeta$. The reduced $\chi^2$ values refer to: training data (blue line), training observables but new continuous (demuxed) trajectories (orange line), test observables and all trajectories (green line). Explanatory comments on the behaviour of the $\chi^2$ are reported in the \textit{Cross validation procedure} section.

\subsubsection{Statistical efficiency}

\begin{figure}
\centering
\includegraphics[width=0.5\textwidth]{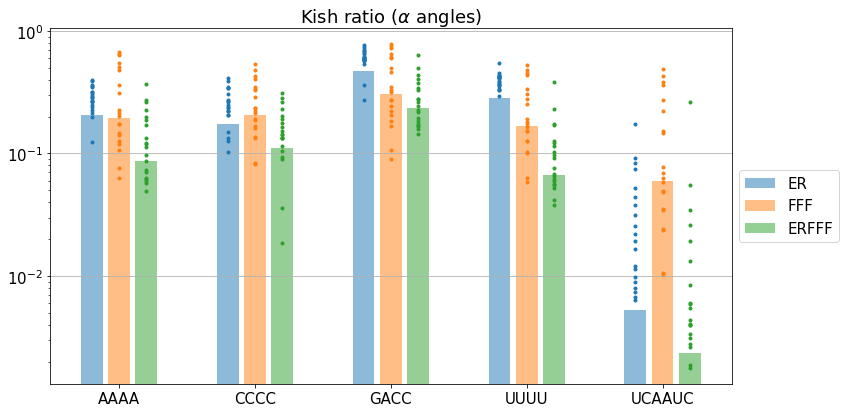}
\captionsetup{justification=raggedright, singlelinecheck=off} %
\caption{Kish ratio (ratio between effective and total number of frames); bars show values from minimization on whole data set, points display values in cross validation values (force field correction on $\alpha$ dihedral angles).}
\label{fig:kish_ratio}
\end{figure}

In Fig. \ref{fig:kish_ratio} we report the Kish ratios $K_{rel}$ for the three methods in the case of force-field correction on $\alpha$ dihedral angles (average over cross validation values). In all the cases the effective number of frames is a significant fraction of the total number of frames.

\begin{figure}
\centering
\includegraphics[width=0.5\textwidth]{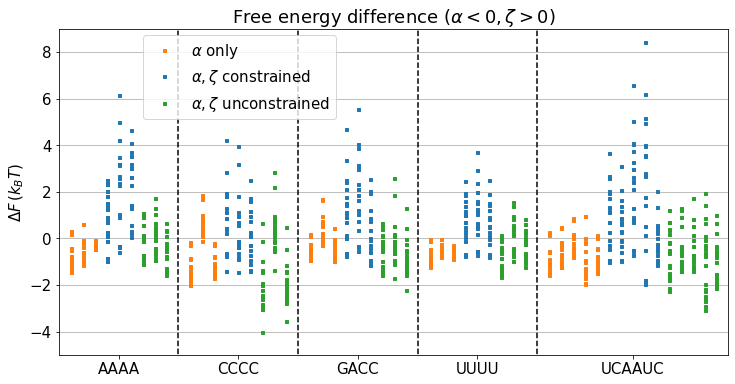}
\captionsetup{justification=raggedright, singlelinecheck=off} %
\caption{Free energy differences for the few-visited region $\alpha<0,\,\zeta>0$ with respect to the original ensembles, for each oligomer and phosphate (comparison of three different force-field correction terms).}
\label{fig:few_visited}
\end{figure}

\subsubsection{Overfitting in low population regions}

In Fig. \ref{fig:few_visited} we report the free-energy differences for the few-visited region $\alpha<0,\,\zeta>0$ between reweighted and original ensembles, for each oligomer and phosphate, resulting from cross-validation. In particular, we focus on the reweighted ensembles corresponding to FFR only, for the cases of sinusoidal corrections on $\alpha$ dihedral angles only, on $\alpha,\zeta$ with same coefficients, on $\alpha,\zeta$ independently. The typical values of dispersion, evaluated as $\sqrt{\frac{1}{N}\sum_{i=1}^N \sigma_i^2}$ with sum over different molecules and phosphates, is $0.49,\, 1.56,\, 0.93\, k_B T$ respectively for the three force-field corrections. This clearly shows the correction on $\alpha$ dihedral angles only is the least sensitive to shifts in the two few-visited regions corresponding to 2nd and 4th quadrants of the $\alpha,\zeta$ plane.

\end{document}